# Predicted disease compositions of human gliomas estimated from multiparametric MRI can predict endothelial proliferation, tumor grade, and overall survival

Emily E Diller[1,2], MS, Sha Cao[3], PhD, Beth Ey[4], MD, Robert Lober[5], MD, PhD, Jason G Parker[1,2], PhD,

[1]Department of Radiology and Imaging Sciences, Indiana University School of Medicine; [2]School of Health Sciences, Purdue University; [3]Department of Biostatistics, Indiana University School of Medicine; [4]Radiology, Dayton Children's Hospital; [5]Neurosurgery, Dayton Children's Hospital

*Corresponding Author Info*: Emily Diller, 950 W. Walnut St., Indianapolis, IN 46202, telephone: (317) 278-9811, dillere@purdue.edu, emdiller@iu.edu

**Abstract:**

*Background and Purpose:* Biopsy is the main determinants of glioma clinical management, but require invasive sampling that fail to detect relevant features because of tumor heterogeneity. The purpose of this study was to evaluate the accuracy of a voxel-wise, multiparametric MRI radiomic method to predict features and develop a minimally invasive method to objectively assess neoplasms.

*Methods:* Multiparametric MRI were registered to T1-weighted gadolinium contrast-enhanced data using a 12 degree-of-freedom affine model. The retrospectively collected MRI data included T1-weighted, T1-weighted gadolinium contrast-enhanced, T2-weighted, fluid attenuated inversion recovery, and multi-b-value diffusion-weighted acquired at 1.5T or 3.0T. Clinical experts provided voxel-wise annotations for five disease states on a subset of patients to establish a training feature vector of 611,930 observations. Then, a k-nearest-neighbor (k-NN) classifier was trained using a 25% hold-out design. The trained k-NN model was applied to 13,018,171 observations from seventeen histologically confirmed glioma patients. Linear regression tested overall survival's (OS) relationship to predicted disease compositions (PDC) and diagnostic age ($\alpha = 0.05$). Canonical discriminant analysis tested if PDC and diagnostic age could differentiate clinical, genetic, and microscopic factors ($\alpha = 0.05$).

*Results:* The model predicted voxel annotation class with a Dice similarity coefficient of 94.34% ± 2.98. Linear combinations of PDCs and diagnostic age predicted OS ($p = 0.008$), grade ($p = 0.014$), and endothelia proliferation ($p = 0.003$); but fell short predicting gene mutations for TP53BP1 and IDH1.

*Conclusions:* This voxel-wise, multi-parametric MRI radiomic strategy holds potential as a non-invasive decision-making aid for clinicians managing patients with glioma.

## 1. Introduction

Every year more than 6 per 100,000 adults in the United States are diagnosed with glioma, the most common malignant tumor of the central nervous system.[1] For clinical purposes, the World Health Organization (WHO) grades gliomas I to IV, based on histologic and molecular features, with the worst survival associated with glioblastoma, WHO grade IV, in which only one half of patients survive one year after diagnosis.[2] Lower grade gliomas (LGG), as defined by The Cancer Genome Atlas (TCGA), include WHO grades II and III gliomas and have a survival range of one to fifteen years.[3] Grading informs various treatment protocols including resection, chemotherapy, radiation therapy, and long-term monitoring. Unfortunately, complete surgical resection may be infeasible, and both epigenetic characteristics and tumor heterogeneity may reduce sensitivity to chemotherapy and radiotherapy.[2]

The current standard for glioma diagnosis is histopathologic evaluation after resection or biopsy, with refinements in tumor classification based on molecular features. However, not all patients are candidates for surgery and, even with advancements in stereotaxic methods, the availability and quality of diagnostic tissue is constrained by procedure time, sampling error, and user interpretation.[5, 6] Therefore, alternative non-invasive methods must be developed to quantitatively investigate glioma phenotypic heterogeneity, which could alter clinical management strategies.

Magnetic resonance imaging (MRI) is a noninvasive imaging modality that provides critical information for glioma detection and diagnosis. Routine protocols usually include the qualitative imaging sequences T1-weighted (T1), gadolinium contrast-enhanced T1 (T1-GD), T2-weighted (T2), and fluid-attenuated inversion recovery (FLAIR).[7, 8] More advanced quantitative MRI sequences, such as diffusion-weighted imaging (DWI) and derived apparent diffusion coefficient (ADC) maps, have predictive potential for glioma differentiation.[9] However, most studies focus on the power of a single MRI sequence, using measurement techniques that neglect heterogeneity.

Radiomics is a recently defined discipline of medical imaging that utilizes quantitative feature extraction and machine-learning models to develop clinically significant predictions.[10] There have been many promising studies using radiomics principles for various cancers, often focused on anatomical locations such as breast, lung, and pancreas, with fewer focused on gliomas.[11-14] The Brain Tumor Segmentation (BraTS), developed by Bakas et al, is a collection of glioma annotations based on T1, T1-GD, T2, and FLAIR MRI data, recently applied to predict overall survival (OS) and progression free survival (PFS) based on radiomic features.[15-17] Inano et al used diffusion tensor imaging (DTI) in a voxel-by-voxel (voxel-wise) method to develop k-means clusters that significantly differentiated WHO grade II from WHO grade III and IV gliomas.[18] Tian et al used multiparametric MRI data from T1, T1-GD, T2, DWI, and arterial spin labeling (ASL) sequences to define texture features that significantly differentiated glioma grades using a more complex and resource intensive support vector machine (SVM) model.[19] However, the utilization of multiparametric MRI with a voxel-wise radiomics method for predicting glioma grade, genetic mutations, and prognosis has not been fully verified.

We hypothesized that a voxel-wise radiomics method using multiparametric MRI data could provide phenotypic classification reflecting general tumor heterogeneity (predicted disease compositions, PDC), with predictive utility for glioma grade and genetic mutations. We tested this with a non-parametric machine learning model employing k-nearest neighbor (k-NN) in a voxel-wise based feature vector across five MRI sequences from a publicly available data set (http://www.iu.edu/~mipl), with phenotype classifications defined by field experts.

## 2. Methods

*2.1 Patient population*

We obtained anonymized medical data for 28 patients with histologic diagnoses of LGG (N = 14) or glioblastoma multiforme (GBM, N = 14) from TCGA, in which WHO grade II and III glial tumors were designated as LGG.[3] We excluded 11 patients: five for incomplete brain coverage, and six for motion or artefact. Therefore, we included 17 patients (10 LGG, 7 GBM; clinical characteristics summarized in **Table 1**).

| | Lower Grade (LGG) | Glioblastoma (GBM) | p-value |
|---|---|---|---|
| **Demographics** | | | |
| Patients | 58.8% (10/17) | 41.2% (7/17) | N/A |
| Age at diagnosis, years (mean ± SD) | 51.88 ±10.67 | 65.72 ±12.91 | 0.029 |
| Gender | | | 0.653 |
| Male | 60.0% (6/10) | 71.4% (5/7) | |
| Female | 40.0% (4/10) | 28.6% (2/7) | |
| PFS, months (mean ± SD) | 20.75 ±13.30 | 4.95 ±7.51 | 0.003* |
| OS, months (mean ± SD) | 34.72 ±15.74 | 15.04 ±1.05 | 0.016* |
| Histologic subtype | Oligoastrocytoma WHO II, 40% (4/10) Oligodendroglioma WHO II, 20% (2/10) Anaplastic oligoastrocytoma WHO III, 20% (2/10) Anaplastic astrocytoma WHO III, 20% (2/10) | Glioblastoma WHO IV, 100% (7/7) | N/A |
| **Genetic Mutation Status (Wild-type : Mutated)** | | | |
| CDKN2A | 9:1 | 3:4 | 0.037 |
| TP53 | 3:7 | 5:2 | 0.104 |
| EGFR | 9:1 | 3:4 | 0.037 |
| NF1 | 7:3 | 6:1 | 0.484 |
| CDKN2B | 10:0 | 3:4 | 0.004* |
| CDK4 | 8:2 | 5:2 | 0.704 |
| TP53BP1 | 6:4 | 7:0 | 0.061 |
| IDH1 | 1:9 | 6:1 | <0.0001* |
| **Histologic Expression Status (Not present : present)** | | | |
| Endothelial proliferation | 10:0 | 1:6 | <0.0001* |
| Palisading necrosis | 10:0 | 4:3 | 0.021 |

**Table 1:** Summary of clinical characteristics with Bonferroni corrected p-values.

*2.2 Imaging Data*

We obtained pre-intervention MRI data for all 17 patients from The Cancer Imaging Archive (TCIA). For each we selected five sequences in the axial plane: (1) T1; (2) T2; (3) FLAIR; (4) DWI; and (5) T1-GD. DWI data were processed into quantitative ADC maps using a custom script that solved the Stejskal-Tanner equation at each voxel.[20] The DWI sequences used two b values (0, and 1000 s/mm$^2$) over two, four, or five directions for ten, one, and six patients, respectively. Due to the nature of the archive, field strength, manufacturer, and coil selection were inconsistent across patients. **Table 2** summarizes the MRI parameters across respective sequences.

| MR Parameter | T1 | T1+ | T2 | FLAIR | ADC |
|---|---|---|---|---|---|
| TR [ms] (0018, 0080) | 2315.97 ±961.99 [500.00, 3116.00] | 2645.97 ±1028.42 [500.00, 3236.34] | 3427.45 ±449.42 [3000.00, 4000.00] | 10002.35 ±0.79 [10002.00, 10004.00] | 10000.00 ±0.00 [10000.00, 10000.02] |
| TE [ms] (0018, 0081) | 7.81 ±2.98 [6.36, 14.00] | 7.81 ±2.98 [6.36, 14.00] | 92.59 ±27.17 [22.00, 104.97] | 130.51 ±11.69 [123.50, 155.00] | 75.45 ±7.40 [71.80, 99.00] |
| Inversion Time [ms] (0018, 0082) | 1084.21 ±184.37 [860.00, 1238.00] | 1211.00 ±101.02 [860.00, 1238.00] | 0.00 ±0.00 [0.00, 0.00] | 2241.18 ±19.65 [2200.00, 2250.00] | 0.00 ±0.00 [0.00, 0.00] |
| Spacing between slices (0018, 0088) | 4.65 ±0.79 [3.00, 5.00] | 2.59 ±0.20 [2.50, 3.00] | 4.65 ±0.79 [3.00, 5.00] | 2.59 ±0.20 [2.50, 3.00] | 4.81 ±0.75 [3.00, 6.00] |
| Acquisition matrix (0018, 1310) | 313.60 ±17.94 [256.00, 320.00] | 313.60 ±17.94 [256.00, 320.00] | 317.87 ±19.00 [256.00, 352.00] | 311.47 ±22.52 [256.00, 320.00] | 128.00 ±0.00 [128.00, 128.00] |
| | 219.73 ±11.26 [192.00, 224.00] | 219.73 ±11.26 [192.00, 224.00] | 219.73 ±11.26 [192.00, 224.00] | 219.73 ±11.26 [192.00, 224.00] | 128.00 ±0.00 [128.00, 128.00] |
| Pixel spacing (0028, 0030) | 0.55 ±0.18 [0.47, 0.94] | 0.55 ±0.18 [0.47, 0.94] | 0.55 ±0.18 [0.47, 0.94] | 0.57 ±0.19 [0.47, 0.94] | 1.00 ±0.23 [0.94, 1.88] |
| Slice Thickness (0018, 0050) | 4.65 ±0.79 [3.00, 5.00] | 2.59 ±0.20 [2.50, 3.00] | 4.65 ±0.79 [3.00, 5.00] | 2.59 ±0.20 [2.50, 3.00] | 4.81 ±0.75 [3.00, 6.00] |
| Number of Averages (0018, 0083) | 1.31 ±0.79 [1.00, 4.00] | 1.31 ±0.79 [1.00, 4.00] | 2.00 ±0.00 [2.00, 2.00] | 1.19 ±0.75 [1.00, 4.00] | 1.00 ±0.00 [1.00, 1.00] |

Parameters constant across sequences:
Field strength 2.90 ±0.39 [1.50, 3.00]
Flip angle 90.00 ±0.00 [90.00, 90.00]
Mean ± standard deviation; [min, max]

**Table 2:** Summary of MR Sequence Settings

*2.3 Image Annotation*

Within our research team, a neurosurgeon and radiologist independently annotated regions of interest on a pre-selected T1-GD slice, after reviewing axial MRI data (T1, T2, FLAIR, ADC, T1-GD) for each patient on a local picture archiving and communication system (PACS). We annotated diseased regions with high confidence (>95%) in categories of (1) pure cyst without necrosis, (2) necrosis, (3) tumor, or (4) edema. For regions with moderate confidence of disease (>50%) but unknown classification, the reader could annotate the area as "suspicious" for disease. We also annotated uninvolved, normal appearing regions in categories of (1) white matter (WM), (2) gray matter (GM), (3) cerebral spinal fluid (CSF), or (4) air.

*2.4 Image Registration and Feature Vector*

We registered each MRI sequence to its respective T1-GD scan in FSL (Analysis Group, FMRIB v5.0, Oxford, UK) using an affine 12-parameter model with a correlation ration cost function and tri-linear interpolation, spatially smoothed the data with a one-mm Gaussian filter, and normalized the qualitative sequences (T1, T1-GD, T2, and FLAIR) to each patient's average normal-appearing WM. From the annotations we created a labeled matrix of feature vectors in which each observation represented one voxel within an annotated region across five MRI sequences. For disease annotations, only voxels with agreement between our neurosurgeon and radiologist were included in the feature vector. The resultant matrix contained 611,930 observations (voxels) from fourteen patients, across five feature vectors (MRI sequences), where each observation belonged to one of nine classes (annotation labels).

*2.5 k-NN Radiomics Algorithm*

We developed a radiomics algorithm based on the k-NN classification model in MATLAB® (R2017a, The MathWorks, Inc.), with model parameters listed in **Table 3**. Based on exploratory methods, the k-NN model demonstrated comparable or higher accuracy with lower computational requirements compared to other parametric and non-parametric machine learning methods. Our model trained on a randomly selected three-fourths of the labeled observations (N = 458,948) and tested on the remaining one-fourth (N = 152,982). The developed radiomics algorithm then predicted the class of each voxel from eleven slices across five MRI contrasts as new observations from seventeen patients, resulting in a total of 13,018,171 class predictions. The predicted class slices included the disease annotation slice, and five slices superior and inferior to the diseased annotation slice. Next, we defined the PDC phenotypes for each classified slice as a distribution of percent suspicious (%Suspicious), percent edema (%Edema), percent tumor (%Tumor), percent cyst (%Cyst), and percent necrosis (%Necrosis).

| Property | Value |
| --- | --- |
| N Neighbors | 10 |
| Distance | Euclidean |
| Include ties | False (0) |
| Distance weight | Equal |
| Break ties | Smallest |
| NS Method | Kdtree |
| Bucket size | 50 |
| Standardize data | True (1) |
| Mu | [105.77, 124.69, 144.49, 67.62, 148.69] |
| Sigma | [134.92, 150.77, 182.78, 74.05, 219.84] |
| Prior probability | [0.0177, 0.0089, 0.0831, 0.0002, 0.0002, 0.1448, 0.0447, 0.5335, 0.1665] |

**Table 3**: Summary of kNN model parameters

*2.6 Statistical Analysis*

We performed statistical analyses with SPSS (IBM, Version 25), with an a priori α significance level of 0.05. We tested the assumption of normality for each continuous variable and applied Box-Cox transformations when necessary. We compared values of age at diagnosis, gender, binary genetic mutation status (CDKN2A, TP53, EGFR, NF1, CDKN2B, CDK4, TP53BP1, IDH1), and binary features of endothelial proliferation (EP) and palisading necrosis (PN) between LGG and GBM using student's t-test. We compared PFS and OS between groups using the log rank test, and tested the relationship between both PFS and OS with PDC phenotypes using linear regression. Although PFS and OS are censored variables, linear regression was appropriately applied because all cohort subjects experienced the same events. We used canonical discriminant analysis with a stepwise Wilk's lambda model that included independent factors with F probabilities < 0.1 and a Bonferroni correction to test differentiation power of one or more disease components on dependent categorical factors with significant prognostic implications. The continuous variables included age at diagnosis and PDC phenotypes (%Suspicious, %Edema, %Tumor, %Cyst, and %Necrosis). The dependent variables included binary categorical factors of tumor grade (TCGA classification of LGG versus GBM), mutation status (wild-type versus mutant) of the abovementioned genes, and the histologic features of EP and PN. Genetic mutation status by patient is shown in **Figure 1**. We computed prior probabilities based on each dependent variable's group size and tested the final model with a cross-validation method.

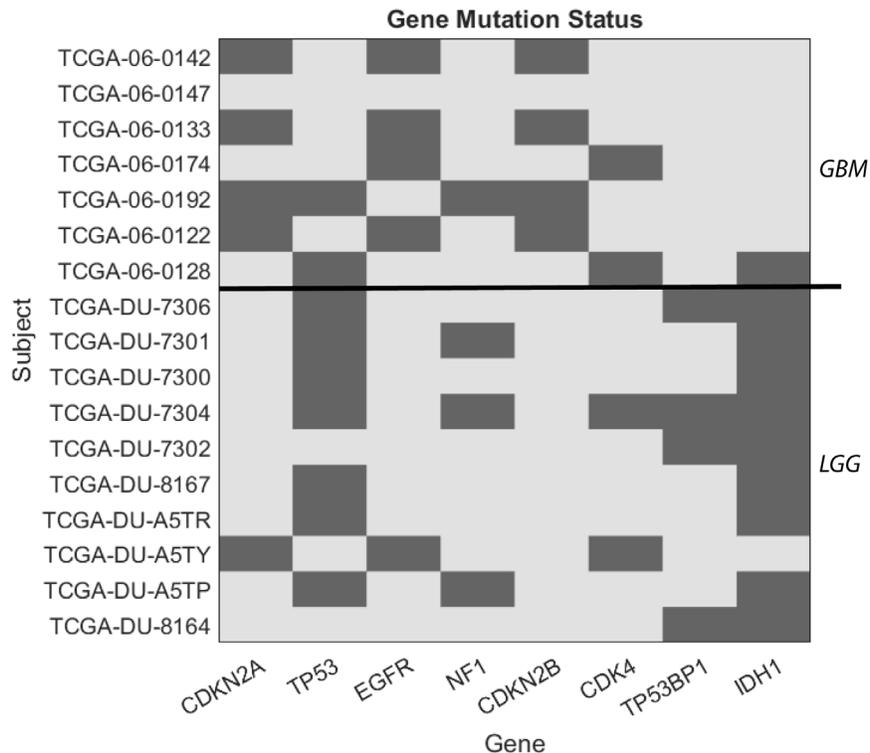

**Figure 1:** Selected genetic mutation status defined as wild-type (light) or mutant (dark) by patient and TCGA grade class. Patients above the bold black line were diagnosed as GBM. Patients below the bold black line were diagnosed as LGG.

## 3. Results

*3.1 k-NN Radiomics Algorithm Performance and Accuracy*

The k-NN model demonstrated 97.0% average accuracy on the training data. On the five diseased tissue classifications, the average accuracy was 95.61 ± 1.48%, with "Suspicious" having the highest performance at 98.4% and "Cyst" the poorest performance at 94.29%. **Figure 2** shows the confusion matrix between the true and predicted classes for the testing data (N = 152,982). The Dice similarity coefficient (DSC) was calculated for each classification (**Table 4**).

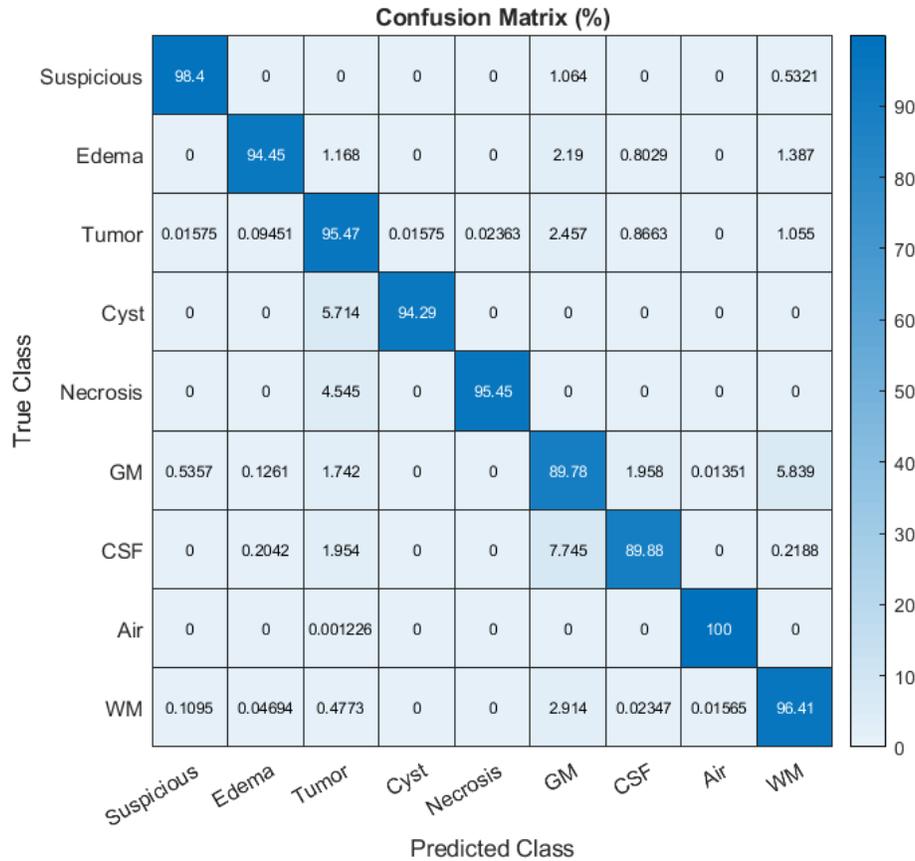

**Figure 2:** Radiomics algorithm k-NN Confusion Matrix. The k-NN model accuracy was tested using a 25% hold-out method. As observed, there is strong main axis agreement between the true and predicted classes. The accuracy for the entire model, including disease and normal tissue classes, was 97.0%. The average accuracy for the diseased classes was 95.61%.

|  | Suspicious | Edema | Tumor | Cyst | Necrosis | GM | CSF | Air | WM |
|---|---|---|---|---|---|---|---|---|---|
| $N_{truth}$ | 2631 | 1370 | 12697 | 35 | 22 | 22212 | 6856 | 81596 | 25563 |
| $N_{predicted}$ | 2589 | 1294 | 12122 | 33 | 21 | 19943 | 6162 | 81595 | 24646 |
| DSC | 96.44 | 94.8 | 95.14 | 94.29 | 91.3 | 91.06 | 90.75 | 100 | 95.36 |
| Accuracy | 98.40 | 94.45 | 95.47 | 94.29 | 95.45 | 89.78 | 89.88 | 100.00 | 96.41 |

**Table 4:** Dice similarity coefficient (DSC) computed for each class, based on the predicted class label and ground truth class label.

The average DSC was 94.35 ± 2.98 across all classes, 94.39 ± 1.90 across disease classes, and 94.29 ± 4.35 across normal tissue classes. Example expert annotations and predictions are shown in **Figure 3**.

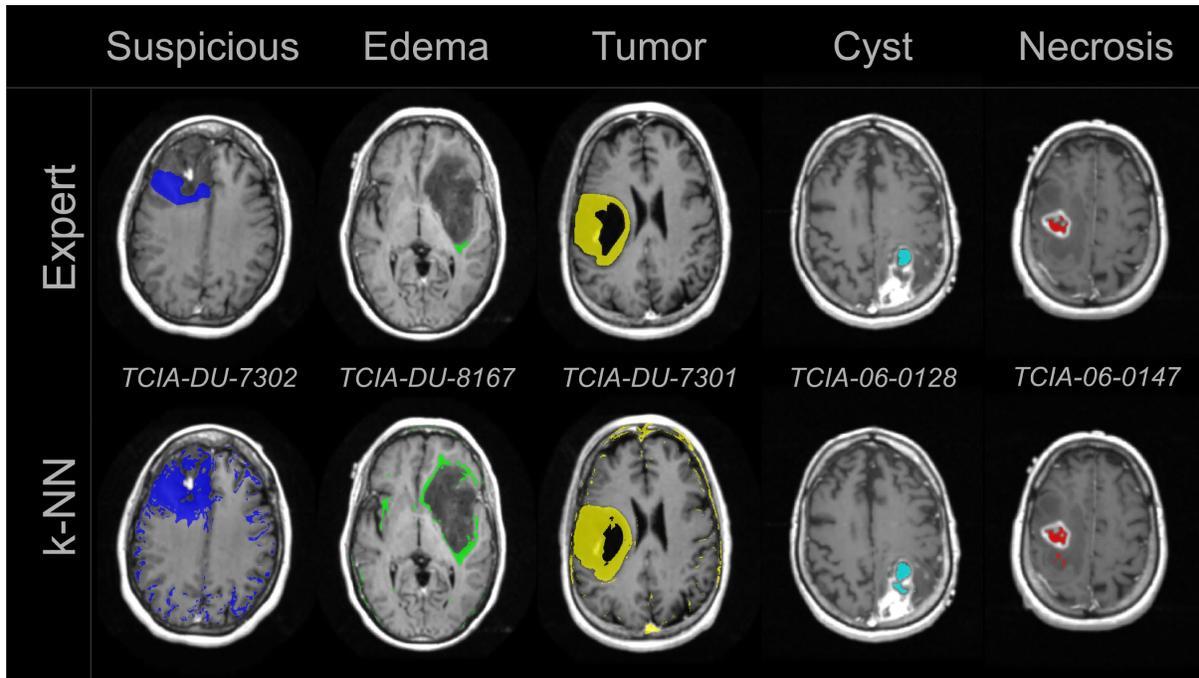

**Figure 3:** Example of expert annotations and k-NN algorithm predictions for diseased classifications suspicious, edema, tumor, cyst, and necrosis.

*3.2 Linear Regression*

In linear regression analysis, the continuous variable %Tumor predicted OS [$F(1, 15) = 7.186$, $p = 0.017$] and accounted for 27.9% of the explained variability in OS. The regression equation was: Predicted OS=7.67-0.037(%Tumor). There was no relationship between PDC phenotypes and PFS.

*3.2 Canonical Discriminant Analysis*

The CDA determined that linear functions of one or more disease components differentiated the binary status of four out of 11 dependent categorical variables (**Table 5**). A linear combination of %Suspicious and age at diagnosis differentiated LGG from GBM with 76.5% accuracy, 85.7% sensitivity, and 70.0% specificity ($p = 0.014$). The discriminant equation to maximally separate discriminant score (DS) by TCGA study class was: DS(TCGA)=0.575(%Suspicious)-0.057(Age at Diagnosis)+2.472. The discriminant equation to maximally separate DS by IDH1 mutational status was: DS(IDH1)=0.040(%Tumor)-2.997. EP status was differentiated with 88.2% classification accuracy, 66.7% sensitivity, and 100% specificity by %Necrosis and Age at Diagnosis ($p=0.003$). The discriminant equation to maximally separate DS by EP status was: DS(EP)=2.479(%Necrosis)+0.066(Age at Diagnosis)-24.253.

|  | TCGA | EP | TP53BP1 | IDH1 |
|---|---|---|---|---|
| Accuracy (%) | 76.5 | 88.2 | 82.4 | 70.6 |
| Sensitivity (%) | 85.7 | 66.7 | 50.0 | 70.0 |
| Specificity (%) | 70.0 | 100 | 92.3 | 71.4 |
| PPV (%) | 66.7 | 100 | 66.7 | 77.8 |
| NPV (%) | 87.5 | 84.6 | 85.7 | 62.5 |
| p value | 0.014* | 0.003* | 0.097 | 0.054 |
| Wilk's λ | 0.544 | 0.442 | 0.827 | 0.774 |
| $\chi^2$ | 8.533 | 11.420 | 2.762 | 3.720 |
| df | 2 | 2 | 1 | 1 |
| R canonical | 0.676 | 0.747 | 0.416 | 0.476 |
| **CDA Model Coefficients** | | | | |
| Suspicious | 0.575 | - | 0.711 | - |
| Edema | - | - | - | - |
| Tumor | - | - | - | 0.040 |
| Necrosis | - | 2.479 | - | - |
| Age at Diagnosis | -0.057 | 0.066 | - | - |
| Constant | 2.472 | -4.253 | -0.989 | -2.997 |
| **CDA Model Discriminant Score ANOVA** | | | | |
| Low state (-) | LGG | Not present | Wild type | Wild type |
| High state (+) | GBM | Present | Mutant | Mutant |
| Mean (-) | 0.711 | -0.791 | -0.239 | 0.630 |
| 95% CI (-) | (0.036, 1.386) | (-1.432, -0.150) | (-0.829, 0.352) | (-0.181, 1.440) |
| Mean (+) | -1.041 | 1.412 | 0.774 | -0.409 |
| 95% CI (+) | (-1.848, -0.234) | (0.544, 2.280) | (-0.291, 1.839) | (-1.087, 0.269) |
| p value | 0.003 | 0.001 | 0.097 | 0.054 |
| Effect Size | 0.457 | 0.558 | 0.173 | 0.226 |
| Observed Power | 0.912 | 0.981 | 0.382 | 0.500 |

**Table 5:** Summary of Canonical Discriminant Analysis statistical findings. Accuracy, sensitivity, specificity, positive predictive value (PPV), and negative predictive value (NPV) are reported in percentage. Low (-) and high (+) state defined for CDA to maximally separate derived discriminant scores (DS). Confidence intervals (CI) were determined for each state from the DS calculated by the respective CDA model coefficients.

## 4. Discussion

Accurate glioma grading is critical for precise therapeutic planning.[21] Histopathologic glioma grading, the diagnostic gold standard, has inherent limitations. Biopsies are often taken from areas of contrast enhancement that may fail to accurately characterize intratumoral heterogeneity and tumoral infiltration.[22, 23] Here we developed a voxel-wise radiomics method using multiparametric MRI data to PDC and their relevance to microscopic and molecular features. The major finding of this study is that PDC from multiparametric MRI data differentiates lower grade gliomas from GBM at the resolution of a single MRI voxel with accuracy, sensitivity, and specificity greater than or equal to 70%.

A unique tool for glioma research is BraTS: a multiparametric MRI data set of histologically confirmed LGG and GBM patients, which includes pre-intervention MRI data (T1, T2, T1Gd, FLAIR) registered to a universal, anatomical template (MNI-152) with manual segmentations for necrosis, edema, and contrast-enhancing tumor.[15] However, previous studies with this have struggled to discriminate LGG from GBM with specificity greater than or equal to 70%.[24] Here we created a model with multiparametric MRI data from glioma patients in the TCGA, yielding strong agreement between true and predicted disease classes with 95.6% and 94.4% average DSC.

The continuous variable %Tumor inversely correlated with OS, which was worse in GBM compared to LGG (p = 0.016), as expected (**Table 1**). Although our CDA model did not find that %Tumor or %Necrosis predicted grade, GBM had higher %Tumor, %Necrosis, and age at diagnosis compared to LGG, which interestingly had greater %Suspicious (p = 0.035, 0.016, 0.049, and 0.015, respectively), a classification defined during expert annotation as regions of moderate confidence for disease and potentially reflecting a more homogeneous abnormal compartment in LGG.[28] Additionally, we found differential power between LGG and GBM by a linear function of %Suspicious, and age at diagnosis. Specifically, with GBM being associated with a decrease in PDC Suspicious and an increase in age. Patients with GBM are generally older at diagnosis than patients with LGG.[25-27]

The most common pattern of confusion occurred when the true disease classes "Cyst" and "Necrosis" were predicted as "Tumor" and "Edema," and when "Tumor" was predicted as "normal-appearing GM". Such labeling confusion is expected, just as inter- and intra-observer error is unavoidable,[21, 22] and we attempted to mitigate this by defining disease class ground truth as voxels of agreement between the independent annotation of two expert clinicians.

Despite the promising results, this study has limitations. First, this retrospective study has a small sample size with variations in MRI protocols, manufacturers, and field strengths across different institutions. Although these variations may help support the generalizability of our findings, a future large-scale study is required to fully assess the generalizability of this model. Second, due to the limited sample size, this study applied machine learning techniques to each voxel as an observation across sequences resulting in a train and test feature vectors with randomly assigned voxels across patients. Since the voxels are randomly assigned to one of two feature vectors, a voxel neighboring a train voxel will be assigned to the test feature vector and could lead to over fitting. In the future, train and test feature vectors distributed by patient instead of random voxel selection should be explored. Third, the disease classification is based on annotation agreement between two expert readers following a common protocol. In future work, annotations should be collected from multiple experts following an established protocol and annotations of high agreement across expert readers should be used for a model.

In conclusion, we have proposed a five-feature, voxel-wise model with five phenotype signature described by PDC that have potential as an imaging biomarker to differentiate prognostic features. Non-invasive methods to reliably classify and differentiate prognostic features is an important development for the advancement of glioma treatment management. The results shown in this work demonstrate the promise and need for future development of computer-aided decision making tools through multiparametric, voxel-wise radiomic algorithms. Utilizing the power of radiomics, gliomas may be non-invasively managed resulting in the advancement of treatment and care.

**Acknowledgements and Disclosure.**

The authors would like to thank the cancer genome atlas (TCGA) and the cancer imaging archive (TCIA) for providing data used in this analysis.